\documentclass[twocolumn,epsf,prb,amsmath,amssymb,showpacs]{revtex4}
\usepackage{graphicx}

\begin{document}
\title{Tunable quantum dots in bilayer graphene}
\author{J. Milton Pereira$^{1,2}$ Jr., P. Vasilopoulos$^3$, and F. M. Peeters$^1$}
\address{$^1$Department of Physics, Universiteit Antwerpen Groenenborgerlaan 171, B-2020 Antwerpen, Belgium\\
$^2$Departamento de F\'{\i}sica, Universidade
Federal do Cear\'a, Fortaleza, Cear\'a, $60455$-$760$, Brazil\\
$^3$Department of Physics, Concordia University, 7141 Sherbrooke
Ouest, Montreal, Quebec, Canada H4B 1R6}

\begin{abstract}
We demonstrate theoretically that quantum dots in bilayers of
graphene can be realized. A position-dependent doping breaks the
equivalence between the upper and lower layer and lifts the
degeneracy of the positive and negative momentum states of the
dot. Numerical results show the simultaneous presence of electron
and hole confined states for certain doping profiles and a
remarkable angular momentum dependence of the quantum dot spectrum
which is in sharp contrast with that for conventional
semiconductor quantum dots. We predict that the optical spectrum
will consist of a series of non-equidistant peaks.
\end{abstract}
 \maketitle

Two-dimensional (2D) carbon crystals, such as single-layer and
bilayer graphene, have been the subject of increasing interest due
to the unusual mechanical and electronic properties, which may
lead to their use in novel nanoelectronic devices. The
relativistic-like properties of carriers in single-layer graphene
\cite{zheng,novo3,novo4,shara,zhang} result from the gapless and
approximately linear electron spectrum near the Fermi energy at
two inequivalent points of the Brillouin zone. The charge carriers
in these structures are described as massless relativistic
fermions and are governed by the Dirac equation. In contrast, for
a symmetric graphene bilayer the spectrum is parabolic at the
vicinity of the $K$ points.

Among the unusual properties of single-layer graphene is the
perfect forward transmission across potential barriers, known as
Klein tunneling \cite{klein,Katsnelson}, which is related to the
absence of a gap in the carrier spectrum. This effect prevents the
electrostatic confinement of charged particles and thus the
realization of quantum dots. Recently, alternative strategies have
been proposed to confine charged particles by using thin
single-layer graphene strips \cite{Peres, Efetov} or non-uniform
magnetic fields \cite{Egger}. Here we propose a novel approach by
considering a bilayer graphene, in which a charge imbalance
between the layers gives rise to a gap in the spectrum, that can
be used to create potential barriers \cite{Falko,Geim}. Such
bilayers of graphene can be obtained, e.g., from a graphite
crystal by micromechanical cleavage \cite{novo2}. A recent report
described the synthesis of bilayer graphene sheets by
graphitization of silicon carbide (SiC) surfaces in which the
equivalence of the two graphene layers is broken by their
interaction with the SiC substrate as well as by doping %of
one of them with potassium atoms \cite{Ohta}.

These recent experimental progresses raise the possibility of
introducing a position-dependent modification of the spectrum at
the Dirac point by changing the potassium density at different
regions of the bilayer graphene sheet or by using  microstructured
gates. In this letter we propose a position-dependent potassium
doping to manipulate the band structure of bilayer graphene in
order to create nanometer-scale quantum structures such as quantum
dots. Semiconductor quantum dots have been intensively
investigated both theoretically and experimentally. Up to now,
only quantum dots on {\it graphite}\cite{Bunch} have been
obtained, with sample thicknesses of several nanometers, i.e., of
the order of $10$ stacked graphene layers. Theoretical
calculations show that the properties of a bilayer graphene can be
significantly distinct from such multilayer systems \cite{Bart}.
In the present work we demonstrate that the electronic states of
such quantum dots can differ significantly from the usual
semiconductor-based dots.

The crystal structure of an undoped bilayer of graphene is that of
two honeycomb sheets of covalent-bond carbon atoms coupled by weak
Van der Waals forces. To each carbon atom corresponds a valence
electron and the structure can be described in terms of four
sublattices, labelled A, B (upper layer) and A$^{\prime}$,
B$^{\prime}$ (lower layer). Considering only nearest-neighbor
hoping, the Hamiltonian of the system in the vicinity of the
${\mathbf K}$ point is given, in the continuum approximation, by
\cite{Snyman}
\begin{equation}
\mathcal{H}=\mathcal{H}_0 +(\Delta U/2)\tau_z,
\end{equation}
with %here
\begin{equation}
\mathcal {H}_0=
\begin{pmatrix}
  U_0 & \pi & t & 0 \\
 \pi^\dagger & U_0 & 0 & 0\\
  t & 0 & U_0 & \pi^\dagger\\
 0 & 0 & \pi & U_0
\end{pmatrix}
,
\end{equation}
where $t \approx 400$ meV is the interlayer coupling term, $\pi =
v_F(p_x + ip_y)$, $\hat{\mathbf p} = (p_x,p_y)$ is the 2D momentum
operator, $v_F \approx 1\times 10^6$ m/s, $U_0 = (U_1+U_2)/2$,
$\Delta U = U_1 - U_2$ and $U_{1}$, $U_{2}$ are the potentials at
the two layers, which reflect the influence of doping on one of
them and/or the interaction with an external electric field, i.e.,
the gating. The operator $\tau_z$ assigns a positive (negative)
sign to the upper (lower) layer labels and is defined as
\begin{figure}
\vspace*{-1cm}
\includegraphics*[height=7cm, width=6cm]{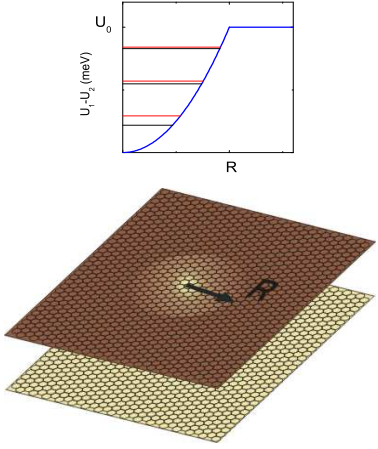}
\vspace*{-0.39cm} \caption{%(Color online).
Lower panel: schematics
of the quantum dot layout, showing the doped graphene layer
(brown) and the quantum dot region (light brown). Upper panel: the
induced parabolic potential and the energy levels of the confined
states.}
\end{figure}
\begin{equation}
\tau_z=
\begin{pmatrix}
  \mathbf{I} & 0  \\
 0 & -\mathbf{I}
\end{pmatrix}
,
\end{equation}
with $\mathbf{I}$ denoting the $2\times 2$ identity matrix. The
eigenstates of Eq. (1) are four-component spinors $\Psi = [\psi_A
\, , \, \psi_B\, , \, \psi_{B'}\, , \, \psi_{A'}]^T$, where
$\psi_{A,B}$ ($\psi_{A',B'}$) are the envelope functions
associated with the probability amplitudes at the respective
sublattice sites of the upper (lower) graphene sheet. One can also
define an {\it inversion} operator $\mathbf X$ as
\begin{equation}
\mathbf{X}=
\begin{pmatrix}
  0 & \mathbf{I}  \\
  \mathbf{I} & 0
\end{pmatrix}
,
\end{equation}
which acts on $\Psi$ by %ex
changing $A$ to $B'$ and $B$ to $A'$. It is seen that for finite
$\Delta U$, the Hamiltonian
(1) does not commute with $\mathbf{X}$;  this %which
results from the lack of inversion symmetry in the system.

For constant potentials $U_1$ and $U_2$, the single-particle
spectrum consists of four bands that correspond to the solutions
of the equation $[s_{F}^2-(E-U_1)^2][
s_{F}^2-(E-U_2)^2]-t^2(E-U_1)(E-U_2)=0$ with $s_{F}=\hbar v_F k$,
given by
\begin{eqnarray}
&&E^+_{\pm}(k) = U_0+(1/2)\big[( t \pm
\Gamma)^2+\Omega\big]^{1/2},\\ &&\cr &&E^-_{\pm}(k) =
U_0-(1/2)\big[( t \pm \Gamma)^2+\Omega
\big]^{1/2},
\end{eqnarray}
where $ \Gamma =\big[t^2+4s_{F}^2 + 4(s_{F}^{2}/t^2)\Delta
U^2\big]^{1/2}$ and $\Omega = \big[1-4s_{F}^2/t^2\big ]\Delta
U^2$. Note that  for $k=0$ the spectrum shows a gap at $k=0$ of
size $E^+_-(0)-E^-_-(0)=|\Delta U|$ and the system becomes a
narrow-gap semiconductor. Figure 2 shows the spectra for  $U_1 =
U_2 = 0$ (dashed red curve), $U_1=200$ meV and $U_2 = 0$ (solid
black), and $U_1=-U_2=100$ meV (dot-dashed blue curve). The
existence of such a gap allows the creation of nanostructures
based on a position-dependent doping, which in turn can create a
position-dependent potential difference between the upper and
lower layers of the structure \cite{Castro2}. Different potentials
$U_1$ and  $U_2$ can be created on defect-free graphene by
controlling the deposition of potassium atoms on the upper layer
or by applying %to it %with %by voltages induced by
gate electrodes \cite{Geim}.

\begin{figure}
\vspace*{-1.5cm} \hspace*{-1cm}
\includegraphics*[height=7cm, width=8cm]{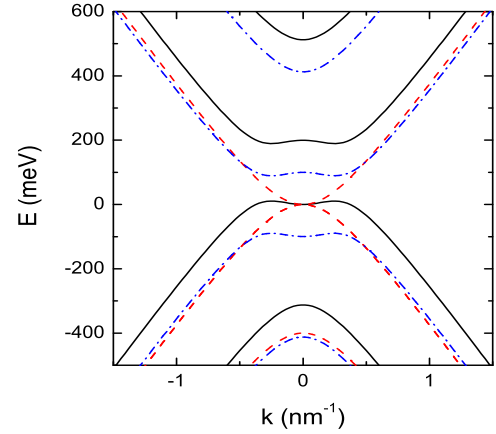}
\vspace*{-0.69cm} \caption{%Color online.
Spectrum of a graphene
bilayer with $U_1 = U_2 = 0$
 (dashed curve), $U_1 = 200$ meV and $U_2 = 0$ (solid curve),  and
$U_1=-U_2 = 100$ meV (dot-dashed curve).}
\end{figure}

The next step is to use a non-uniform doping, or %alternatively
microstructured gate electrodes, that will lead to a
position-dependent potential difference between the layers. In the
experimental system studied
 in Ref. \onlinecite{Ohta} there was a potential
difference $\Delta U$ between the graphene layers even in the absence of
doping, due to their interaction with the SiC substrate. In that
case the potentials were equalized when the number of doping
electrons per unit cell was $\gamma_0 = 0.0125$. For the sake of
simplicity we assume a constant interlayer hopping term $t$ though %which
in reality $t$ may % also
depend on the doping level. In the present case a small change of
the value of $t$ will not affect significantly the results for low
energies. The experimental results show that  $\Delta U$ scales
approximately linearly with doping over a wide range of potassium
concentration levels.
%From the data of Ref. \onlinecite{berger} we
%could fit
%\begin{equation}
%\Delta U = U_{1}-U_{2}=16\big[\gamma({\mathbf r}) - \gamma_0\big],
%\end{equation}
%where the
%number of doping electrons per unit cell $\gamma({\mathbf r})$ is a smooth function %denoting as a function
%of %position
%${\mathbf r}$ and  $\Delta U$ is in eV.

A position-dependent, circular-symmetric profile %of the type
$\gamma({\mathbf r})=\gamma(\rho)$ % leads to %allows for
gives rise to a %the creation of a
2D circular-symmetric quantum dot of radius $R$ on a  bilayer
graphene, shown schematically in Fig. 1, with the doping level
ranging from $\gamma_0$ at $\rho = 0$ to a maximum value
$\gamma_M$ for $\rho
> R$. In that case the four-component spinors $\Psi$ are given by%solutions are of the type
\cite{Vincenzo}
\begin{equation}
\Psi(\rho,\theta)=
\begin{pmatrix}
  \phi_A(\rho)e^{im\theta}  \\
  \ \\
  i\phi_B(\rho)e^{i(m-1)\theta}\\
\  \\
  \phi_{B'}(\rho)e^{im\theta}\\
\ \\
  i\phi_{A'}(\rho)e^{i(m+1)\theta}
\end{pmatrix}
,
\end{equation}
where $m$ is an integer. % and
$\Psi$ are eigenstates of the operator
\begin{equation}
J_z = L_z + \frac{\hbar}{2}\tau_z -\frac{\hbar}{2}
\begin{pmatrix}
  \sigma_z & 0  \\
  0 & -\sigma_z
\end{pmatrix}
\end{equation}
with eigenvalue $m$;  $L_z$ is the angular momentum operator and
$\sigma_z$  the Pauli matrix. This represents the total angular
momentum of the carrier %, includes the pseudospin, the Berry geometric phase,
and the layer index operator, which is associated
with the behavior of the system under inversion.

The potential difference between the layers is $\Delta U =
U(\rho)$ and subsequently the equations for the radial part of the
spinor functions are
\begin{eqnarray}
&&\big[ d/d \zeta + m/\zeta\big]\phi_A =
\Omega_1\phi_B,\\ &&\cr &&\big[d/d
\zeta -(m-1)/\zeta\big]\phi_B = -\Omega_1\phi_A + t'\phi_{B'},\\
&&\cr && \big[ d/d \zeta +
(m+1)/\zeta \big]\phi_{A'} = -\Omega_2\phi_{B'}+t'\phi_A,\\
&&\cr &&\big[d/d \zeta -m/\zeta\big]\phi_{B'}
= \Omega_2\phi_{A'},
\end{eqnarray}
where $\zeta=\rho/R$, $\Omega_{1,2} = \epsilon - u_{1,2},$
$u_{1,2}=U_{1,2}R/\hbar v_F$, $\epsilon = ER/\hbar v_F$,  and
$t'=t R/\hbar v_F$. The $m \pm 1$ that appear in Eqs. (10) and
(11) are a direct consequence of the presence of a geometric phase
in the system (i.e. a full rotation of the system introduces a
phase difference of $2\pi$ between the sublattice sites $A'$ and
B). In addition, the large magnitude of the layer coupling term
$t'$, in comparison with the other energy terms, implies that the
$\phi_{B}$ and $\phi_{A'}$ must have significantly higher
amplitudes than $\phi_{A}$ and $\phi_{B'}$. These two facts have
important consequences for the energy spectrum and the shape of
the resulting probability density functions, as we show below.

\begin{figure}
\vspace*{-1cm}
\includegraphics*[height=6.5cm, width=8.5cm]{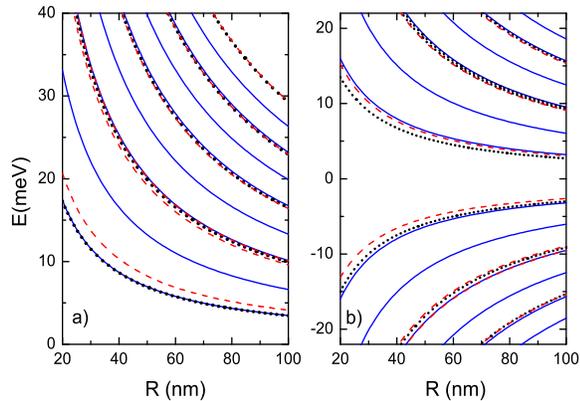}
\vspace*{-0.5cm} \caption{Lowest energy states of a graphene
bilayer quantum dot for $m=0$ (blue solid line), $m=1$ (black
dotted line) and $m=-1$ (red dashed), for (a) $U_M = 50$ meV and
$U_2 = 0$ and (b) $U_1 = -U_2$ and $U_M = 100$ meV.}
\end{figure}
\begin{figure}
\vspace*{-1.2cm} \hspace*{-1cm}
\includegraphics*[height=8.2cm, width=8.5cm]{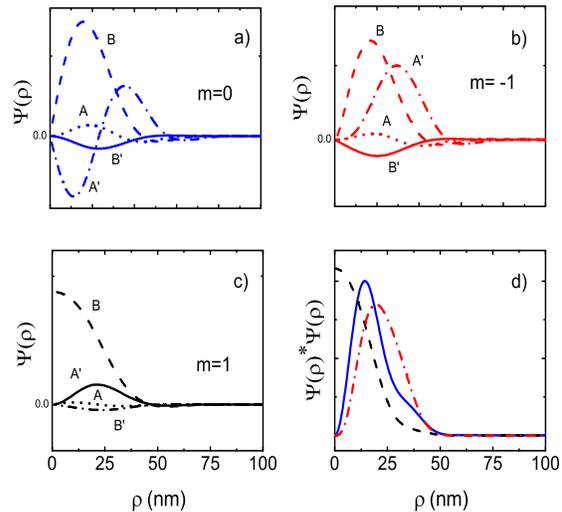}
\vspace*{-0.69cm}
\caption{%Color online.
Panels (a)-(c): Radial
dependence of the spinor functions for a parabolic quantum dot on
a doped bilayer graphene %sheet
for $m=0$, and $m=\pm 1$. Panel (d): probability density for each
case. The parameters are  $R=50$ nm,  $U_1=50$ meV and $U_2=0$.}
\end{figure}

For the numerical calculations we took a truncated parabolic
profile
%\begin{equation}
$\Delta U=U_M\zeta^2$,
%\end{equation}
for $\zeta < 1$ and $\Delta U=U_M$ otherwise. The energies of the
lowest energy bound states are shown in Fig. 3 as function of the
radius of the dot. Figure 3(a) shows the case $U_2 = 0$, $U_M=50$
meV and Fig. 3(b) the case $U_1=-U_2$, $U_M = 100$ meV.
%In the
%case (a), the position-dependent confining potential acts only on
%the electronic states, with the holes states being free, whereas
%in (b) both electrons and holes can be confined.
In both cases the spectrum is proportional to the inverse of the
dot radius and each set of energy levels are approximately equally
spaced for fixed $m$. The results correspond to the lowest energy
levels for $m=0$ (blue solid lines), $m=-1$ (red dashed), and
$m=+1$ (black dotted lines). An important difference between (a)
and (b) is the absence of hole bound states in (a), since there is
an overlap between the hole band states in the well and the
corresponding band in the barriers, whereas in (b) both electron
and hole confined states are present. We found the remarkable
result that the ground state is almost degenerate for $m=0$ and
$m=\pm 1$ in both cases (a) and (b). This turns out to be a
consequence of the geometric phase which shifts the angular
momentum labels in Eqs. (10) and (11). In comparison to the
results for semiconductor quantum dots we find that in a graphene
bilayer there is a peculiar energy shift between states with
positive and negative angular momentum even in the absence of an
external magnetic field. This is a consequence of the lack of
inversion symmetry in the system when $U_1 \neq U_2$. The shift is
larger for the low-lying states, with the $+m$ and $-m$ states
becoming nearly degenerate as the energy increases. A similar,
albeit weaker, shift is found for $|m|>1$.

\begin{figure}
\vspace*{-0.95cm} \hspace*{-1cm}
\includegraphics*[height=8.0cm, width=7.5cm]{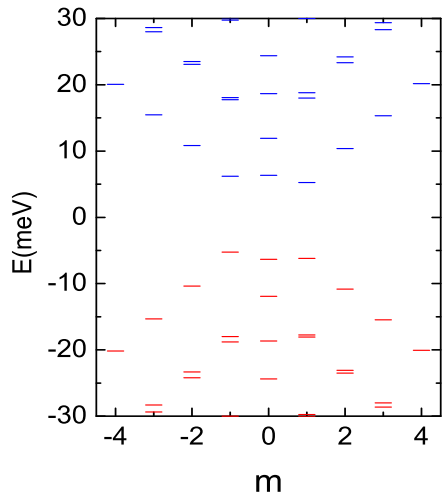}
\vspace*{-0.8cm} \caption{Energy spectrum as a function of angular
momentum label $m$ for $R=50$ nm, and $U_1=-U_2=50$ meV.}
\end{figure}
The radial part of the spinor functions for some %of the
lowest energy states is shown in Fig. 4 for $m=0$ (a), $m=-1$ (b),
and $m=1$ (c). Panel (d) shows the probability density ($\Psi^*
\Psi$) for $m=0$ (solid black curve), $m=-1$ (blue dot-dashed
curve) and $m=1$ (red dashed curve). All results were obtained
%fora truncated parabolic quantum dot with
with $U_1 = 50$ meV, $R=50$ nm, and $U_2=0$. In contrast to a
semiconductor quantum dot, the
present results %in the graphene case
show that the probability density %is
for $m=0$ has a maximum at $\rho \neq 0$. This is due to the fact
that, for $m=0$, the high-amplitude spinor components (i.e.
$\phi_{B}$ and $\phi_{A'}$) vanish at the origin.
%being a direct consequence of the existence of a geometric phase.
%in the system.
In contrast, for $m=1$, which is here the ground state, one of the
high-amplitude spinor functions is allowed to assume non-zero
values at the origin. In other words, for $m=1$, an electron
encircling the dot adds an angular momentum phase of $2\pi$ which
can be cancelled by a Berry's phase of $-2\pi$. The shift between
the low-lying energy levels of positive and negative momentum
labels is evident in Fig. 5, which shows results for a quantum dot
with $U_2=-U_1$ and $R=50$ nm. The results is this case show that
the energy eigenvalues are related by $E(m)=-E(-m)$. This is
consistent with Eqs. (9)-(12) and in sharp contrast with the
relation $E(m)=E(-m)$ satisfied in a usual quantum dot with
parabolic confinement.
%%%%%%%%%%%%%%%%%%%%%%%%%%%%%%%%%%%%%%%%%%%%%%%%%%%%%%%%%%%%%%%%%%%%%%%%
A similar behavior was observed for other potential profiles, such
as the square quantum dot and the tangent hyperbolic potential.
%%%%%%%%%%%%%%%%%%%%%%%%%%%%%%%%%%%%%%%%%%%%%%%%%%%%%%%%%%%%%%%%%%%%%%%%

Figure 6 shows the oscillator strengths for the allowed
transitions $(m,n)\rightarrow (m\pm 1,n')$ between the three
lowest energy states (i.e. $m=0$ and $\pm 1$) within the energy
window of $0 < \Delta E < 9$ meV, in a quantum dot of radius
$R=50$ nm with $U_1 = -U_2 = 50$ meV. Since the energy levels for
different $m$ are not equally spaced, the results show the matrix
elements at several values of $\Delta E$. This differs
considerably from the case of the parabolic quantum dots in
semiconductors, in which the transitions occur at a single energy.
These transitions can be probed by means of far infrared
spectroscopy measurements.

%The matrix elements in this case differ considerably from the case
%of the parabolic quantum dots in semiconductors, in which the
%transitions occur at a single energy, whereas in the present case
%the energy levels are not equally spaced.

\begin{figure}
\vspace*{-1cm} %\hspace*{-1cm}
\includegraphics*[height=7.3cm, width=8.0cm]{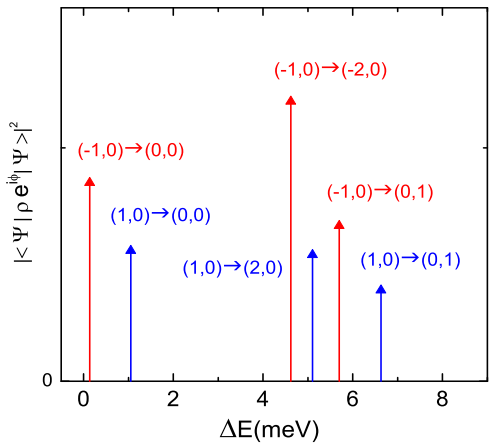}
\vspace*{-0.69cm} \caption{%(Color online)
Oscillator strengths for
transitions between the lowest energy levels $(m,n)$ in quantum
dot of radius $R=50$ nm with $U_1 = -U_2 = 50$ meV.}
\end{figure}

In summary, we proposed a novel approach to electrically confine
charge carriers in 2D quantum dots in bilayer graphene. The
quantum dots are created through a position-dependent doping which
breaks the equivalence between the upper and lower layers. A
further tuning of the quantum dot can be realized through the use
of gates. The numerical results show that the degeneracy of the
positive and negative momentum states of the dot is lifted even in
the absence of an external magnetic field. This result differs
fundamentally  from that for conventional semiconductor quantum
dots and arises due to the lack of inversion symmetry caused by
the doping. The system can be realized experimentally by a
suitable choice of doping levels or with the application of
electric fields.

{\it Acknowledgements}
 This work was supported by the Brazilian
Council for Research (CNPq), BOF-UA, the Flemish Science
Foundation (FWO-Vl), the Belgian Science Policy (IUAP) and the
Canadian NSERC Grant No. OGP0121756.

\end{document}